\documentclass[aps,preprint,showpacs,preprintnumbers,amsmath,amssymb]{revtex4}
\usepackage{amsmath,mathrsfs,amsbsy,color,graphicx,bm,amsthm,amsfonts}
\usepackage{units}
\usepackage{bbm}
\usepackage{times}
\usepackage{dcolumn}
\usepackage{mathrsfs}
\usepackage{amsmath,amssymb,epsfig}
\begin{document}

\title{Asymmetric quantum steering harvested near a Lorentz-violating BTZ black hole}
\author{Si-Yu Liu$^1$, Xin-Ze Song$^1$, Xiang-Yue Yu$^1$, Wentao Liu$^{2,3}$\footnote{wentaoliu@hunnu.edu.cn (corresponding author)}, Xiao-Li Huang$^1$, Shu-Min Wu$^1$\footnote{smwu@lnnu.edu.cn (corresponding author)}}
\affiliation{$^1$  Department of Physics, Liaoning Normal University, Dalian 116029, China \\
$^2$ Lanzhou Center for Theoretical Physics, Key Laboratory of
 Theoretical Physics of Gansu Province, Key Laboratory of Quantum Theory and Applications of MoE, Gansu Provincial Research Center for Basic Disciplines of Quantum Physics, Lanzhou University, Lanzhou 730000, China \\
 $^3$ Institute of Theoretical Physics $ \& $ Research Center of Gravitation, Lanzhou University, Lanzhou 730000, China
}


\begin{abstract}
We investigate the harvesting of quantum steering and its directional asymmetry between two Unruh-DeWitt detectors in a Lorentz-violating BTZ black hole spacetime. Since the detectors are located at different radial positions outside the black hole, they experience inequivalent local environments induced by gravitational redshift, causing Alice to undergo stronger effective thermal noise than Bob. Remarkably, we uncover a counterintuitive phenomenon in which the detector subjected to a higher effective temperature exhibits stronger steerability than the other one, revealing a nontrivial inversion of thermal intuition in curved spacetime.
Furthermore, quantum steering survives only within a finite window of detector energy gaps and reaches its maximum within an optimal regime. We find that Lorentz violation suppresses steering most strongly near this optimal energy gap, indicating an enhanced sensitivity of maximal correlation extraction to symmetry breaking effects. Our results demonstrate that Lorentz violation acts as a geometric constraint on the quantum information capacity of spacetime, simultaneously restricting both the strength and the directionality of quantum correlations.
\end{abstract}

\vspace*{0.5cm}
 \pacs{04.70.Dy, 03.65.Ud,04.62.+v }
\maketitle
\section{Introduction}
Einstein-Podolsky-Rosen (EPR) steering is a form of quantum correlation that captures the ability of one party to nonlocally affect the state of another through local measurements \cite{A5}. Originally introduced by Schr\"{o}dinger and later formalized in an operational framework for mixed states \cite{A6,A7}, steering occupies an intermediate position in the hierarchy of quantum correlations, lying between entanglement and Bell nonlocality \cite{A5,A8}. A defining feature of steering is its intrinsic directionality: the ability of one observer to steer another does not, in general, imply the reverse. This asymmetry can become extreme in the form of one-way steering, where steering is possible only in a single direction while being completely absent in the opposite one  \cite{A9,A10,A11,A12,A13}. Such behavior highlights the sensitivity of steering to imbalances in local environments and measurement settings, making it a particularly powerful probe of asymmetric quantum correlations. These properties also underpin its advantages in one-sided device-independent protocols, including quantum communication, metrology, and networked quantum information processing \cite{A14,A15,A16,A17,A18,A19,A20}.

Relativistic quantum information is an interdisciplinary field at the interface of quantum information theory, quantum field theory, and general relativity. It focuses both on how relativistic effects influence quantum resources and information-processing tasks \cite{SDF1,SDF2,SDF3,SDF4,SDF5,SDF6,SDF7,SDF8,SDF9,SDF10,SDF11,SDF12,SDF13,SDF14,SDF15,SDF16,SDF17,SDF18,SDF19,SDF20,SDF21,SDF22,SDF23,SDF24,SDF25,SDF26,SDF27,SDF28,SDF29,SDF30,SDF31,SDF32,SDF33,SDF34,SDF35,SDF36,SDF37,SDF38,SDF39,SDF40}, and on the use of quantum technologies to simulate gravitational phenomena and probe the structure of spacetime \cite{SDF41,SDF42,SDF43,SDF44,SDF45,SDF46,SDF47,SDF48,SDF49,SDF50}. In the relativistic quantum field setting, the vacuum is far from trivial: beyond local fluctuations, it encodes long-range quantum correlations. These correlations can be accessed in an operational manner using localized probes, typically described by Unruh-DeWitt detectors, which interact locally with the field along prescribed worldlines \cite{QET1,QET2,QET3}. Through such interactions, two detectors that are initially uncorrelated can acquire correlations by effectively sampling the underlying vacuum structure. This process, commonly referred to as entanglement harvesting, offers a direct means to interrogate the nonlocal features of quantum field states, even when the detectors remain spacelike separated. The Unruh-DeWitt framework, owing to its minimal structure and its capability to capture key aspects of light-matter interactions, has emerged as the canonical model for analyzing these phenomena.  A substantial body of work has explored harvesting protocols across diverse physical scenarios, including curved backgrounds, accelerated motion, and the presence of reflecting boundaries \cite{SDF51,SDF52,SDF53,SDF54,SDF55,SDF56,SDF57,SDF58,SDF59,SDF60,SDF61,SDF62,SDF63,SDF64,SDF65,SDF66,SDF67,SDF68,SDF69,SDF70,SDF71,SDF72,SDF73,SDF74,SDF75,SDF76,SDF77,SDF78}. In parallel, ongoing developments in quantum simulation have opened promising avenues for experimental implementation, with candidate platforms ranging from superconducting circuits to ultracold atomic systems and Bose-Einstein condensates \cite{SY1,SY2,SY3}.

Lorentz invariance is a fundamental symmetry underlying quantum field theory.
However, various approaches to quantum gravity, together with indications from high-energy astrophysical observations, have motivated the possibility that this symmetry may be violated at fundamental scales  \cite{SY4}.
In phenomenological descriptions, such effects are typically captured within effective field theory and are expected to appear as low-energy remnants of high-energy symmetry breaking \cite{SY5,SY6}.
A particularly simple realization is provided by bumblebee gravity, in which a vector field acquires a nonvanishing vacuum expectation value, thereby inducing spontaneous Lorentz symmetry breaking and modifying the geometric structure of spacetime \cite{SY7,SY8,SY9,BB1,BB2,BB21,BB22}.
This framework has recently attracted considerable attention in black hole physics \cite{BB3,BB4,B7,BB5,BB6,BB7,BB8,BB9,BB10}.
Within this setting, Lorentz violation modifies the geometry in which the scalar field propagates and thereby affects the field correlations sampled by the Unruh-DeWitt detectors.
In curved spacetime, and especially in black hole backgrounds, detectors located at different radial positions experience inequivalent local conditions, resulting in an intrinsic asymmetry.
While entanglement captures symmetric quantum correlations, quantum steering is inherently directional and thus more sensitive to such environmental inequivalence, including differences in gravitational redshift and local thermal noise.
Near the black hole horizon, where strong gravitational effects and potentially enhanced Lorentz-violating contributions coexist, this asymmetry is expected to be further amplified.
Therefore, black hole spacetimes provide a natural arena to explore how Lorentz violation influences the generation, distribution, and directionality of quantum steering.

Building on the above considerations, we investigate the emergence of directional quantum steering in a Lorentz-violating BTZ black hole background. We consider two identical, static Unruh-DeWitt detectors placed at different radial positions outside the horizon, such that the asymmetry of their local environments naturally induces an asymmetry in steerability. By deriving analytical expressions for steering in both directions, we identify how gravitational redshift and unequal thermal noise give rise to directional imbalance, thereby linking the origin of steering asymmetry directly to spacetime-induced inequivalence.  Within this framework, we further analyze how Lorentz violation reshapes this asymmetry and the overall harvesting process. We show that the extractable steering and its directional imbalance are jointly controlled by detector separation, radial position, and energy gap, while Lorentz violation systematically suppresses both the strength and asymmetry of steering and restricts the parameter regime in which it can persist. These results demonstrate that directional quantum steering provides a sensitive probe of spacetime-induced asymmetry beyond what can be inferred from symmetric quantum correlations.

The paper is organized as follows. In Sec. II, we introduce the quantification of quantum steering for X state.  Sec. III presents the Unruh-DeWitt detector model in the Lorentz-violating BTZ spacetime. In Sec. IV, we analyze the harvesting of quantum steering and its directional asymmetry between two detectors in the bumblebee-induced Lorentz-violating geometry. Finally, Sec. V contains our conclusions and outlook.

\section{Quantification of quantum steering for X state}
In this section, we briefly review the definition and quantification of quantum steering for bipartite systems. We consider a two-qubit system shared between Alice and Bob, described by an X-type density matrix $\rho_{AB}$ of the form
\begin{eqnarray}\label{S21}
\rho_{AB} = \begin{pmatrix}
\rho_{11} & 0 & 0 & \rho_{14} \\
0 & \rho_{22} & \rho_{23} & 0 \\
0 & \rho_{32} & \rho_{33} & 0 \\
\rho_{41} & 0 & 0 & \rho_{44}
\end{pmatrix},
\end{eqnarray}
where the matrix elements satisfy $\rho_{ij}^*=\rho_{ji}$ and $\sum_i \rho_{ii}=1$.
Since quantum steering is directional, we first focus on the steerability from Bob to Alice ($B\rightarrow A$), corresponding to the scenario in which Bob performs local measurements that remotely prepare ensembles of states for Alice. According to the steering criterion introduced in Refs.~\cite{B2,B6}, the state $\rho_{AB}$ is steerable from Bob to Alice if and only if a corresponding auxiliary state $\varsigma_{AB}$ is entangled. This construction maps the steering problem onto an equivalent entanglement detection problem, thereby allowing one to exploit well-established entanglement criteria. The auxiliary state $\varsigma_{AB}$ is defined as
\begin{eqnarray}\label{S23}
\varsigma_{AB} = \frac{\rho_{AB}}{\sqrt{3}} + \frac{3 - \sqrt{3}}{3} \left( \rho_A \otimes \frac{I}{2} \right),
\end{eqnarray}
where $\rho_A=\mathrm{Tr}_B(\rho_{AB})$ is the reduced density matrix of Alice and $I$ denotes the $2\times2$ identity operator. Physically, this construction incorporates both the original bipartite correlations and a locally classical contribution, such that the entanglement of $\varsigma_{AB}$ faithfully witnesses the presence of steering in the original state.
Substituting Eq.(\ref{S21}) into Eq.(\ref{S23}), the explicit matrix form of $\varsigma_{AB}$ reads
\begin{eqnarray}
\varsigma_{AB} = \begin{pmatrix}
\frac{\sqrt{3}}{3}\rho_{11} + e & 0 & 0 & \frac{\sqrt{3}}{3}\rho_{14} \\
0 & \frac{\sqrt{3}}{3}\rho_{22} + e & \frac{\sqrt{3}}{3}\rho_{23} & 0 \\
0 & \frac{\sqrt{3}}{3}\rho_{32} & \frac{\sqrt{3}}{3}\rho_{33} + g & 0 \\
\frac{\sqrt{3}}{3}\rho_{41} & 0 & 0 & \frac{\sqrt{3}}{3}\rho_{44} + g
\end{pmatrix},
\end{eqnarray}
with
\begin{equation}
e=\frac{3-\sqrt{3}}{6}(\rho_{11}+\rho_{22}), \qquad
g=\frac{3-\sqrt{3}}{6}(\rho_{33}+\rho_{44}).
\end{equation}
The entanglement of $\varsigma_{AB}$ can be analytically determined due to its X-state structure. In particular, $\varsigma_{AB}$ is entangled if at least one of the following inequalities is satisfied:
\begin{align}
|\rho_{14}| &> \sqrt{Q_a - Q_b}, \\
|\rho_{23}| &> \sqrt{Q_c - Q_b},
\end{align}
where
\begin{align}
Q_a &= \frac{2 - \sqrt{3}}{2} \rho_{11}\rho_{44} + \frac{2 + \sqrt{3}}{2} \rho_{22}\rho_{33} + \frac{1}{4}(\rho_{11} + \rho_{44})(\rho_{22} + \rho_{33}),  \notag \\
Q_b &= \frac{1}{4}(\rho_{11} - \rho_{44})(\rho_{22} - \rho_{33}), \notag \\
Q_c &= \frac{2 + \sqrt{3}}{2} \rho_{11}\rho_{44} + \frac{2 - \sqrt{3}}{2} \rho_{22}\rho_{33} + \frac{1}{4}(\rho_{11} + \rho_{44})(\rho_{22} + \rho_{33}).
\end{align}
Based on these conditions, the degree of steerability from Bob to Alice can be quantified by
\begin{eqnarray}\label{S25}
S^{B\rightarrow A}
=
\max\Big\{
0,\,
|\rho_{14}|-\sqrt{Q_a-Q_b},\,
|\rho_{23}|-\sqrt{Q_c-Q_b}
\Big\}.
\end{eqnarray}
This quantity vanishes for unsteerable states and increases monotonically with the violation of the corresponding steering inequalities.

Owing to the intrinsic asymmetry of quantum steering, steerability from Alice to Bob ($A\rightarrow B$) must be analyzed independently. In this case, one introduces the auxiliary state
\begin{eqnarray}\label{S24}
\varsigma_{BA}
=
\frac{\rho_{AB}}{\sqrt{3}}
+
\frac{3-\sqrt{3}}{3}
\left(
\frac{I}{2}\otimes\rho_B
\right),
\end{eqnarray}
where $\rho_B=\mathrm{Tr}_A(\rho_{AB})$ is Bob’s reduced density matrix. The state $\rho_{AB}$ is steerable from Alice to Bob if $\varsigma_{BA}$ is entangled, which leads to the conditions
\begin{equation}
|\rho_{14}|>\sqrt{Q_a+Q_b}
\quad\text{or}\quad
|\rho_{23}|>\sqrt{Q_c+Q_b}.
\end{equation}
Accordingly, the steerability from Alice to Bob is quantified as
\begin{eqnarray}\label{S26}
S^{A\rightarrow B}
=
\max\Big\{
0,\,
|\rho_{14}|-\sqrt{Q_a+Q_b},\,
|\rho_{23}|-\sqrt{Q_c+Q_b}
\Big\}.
\end{eqnarray}
The explicit dependence of $S^{B\rightarrow A}$ and $S^{A\rightarrow B}$ on different combinations of matrix elements highlights the fundamental asymmetry of quantum steering, which plays a crucial role in relativistic and gravitational settings.

\section{Unruh-DeWitt detectors in Lorentz-Violating spacetime}
We consider Unruh-DeWitt detectors modeled as two-level quantum systems with energy gap $\Omega$. Specifically, we introduce two identical detectors, labeled by $D \in \{A, B\}$, which interact locally with a scalar quantum field $\phi[\textbf{x}_D(\tau)]$ along their respective worldlines. The interaction Hamiltonian in the interaction picture is given by
\begin{equation}\label{S10}
H_D(t) = \lambda \chi(t) \left( e^{i\Omega_D \tau} \sigma^+ + e^{-i\Omega_D \tau} \sigma^- \right) \otimes \phi[\textbf{x}_D(\tau)],
\end{equation}
where $\sigma^+ = |1_D\rangle\langle0_D|$ and $\sigma^- = |0_D\rangle\langle1_D|$ are the raising and lowering operators acting on the detector Hilbert space. The function $\chi(\tau) = \exp[(\tau - \tau_0)^2 / \sigma^2]$  is a Gaussian switching function that controls the duration of the interaction,  $\lambda$  denotes the coupling strength,  $\Omega_D$ is the energy gap of the detector, and $\textbf{x}_D(\tau)$ represents the spacetime trajectory of detector $D$, parametrized by its proper time $\tau$.

Here the bumblebee and scalar fields play different roles. The bumblebee field is treated as a classical background whose vacuum expectation value modifies the BTZ geometry, whereas the scalar field serves as the quantum probe sampled by the detectors. The modified geometry changes the scalar-field Wightman function and hence the detector response, allowing the detectors to capture the geometric imprint of Lorentz symmetry breaking without coupling directly to the bumblebee field. We use the scalar Unruh--DeWitt interaction because it provides a standard framework for isolating this geometric effect and comparing with previous harvesting studies in BTZ spacetime. A direct detector--bumblebee coupling would require quantizing the bumblebee fluctuations and specifying a detector--vector-field interaction, which is beyond the scope of this work.

We assume that both detectors are initially prepared in their ground states $|0\rangle_A|0\rangle_B$, while the scalar field is in the vacuum state $|0\rangle_M$. The initial state of the total system is thus
\begin{equation}
|\Psi_i\rangle = |0\rangle_A|0\rangle_B|0\rangle_M.
\end{equation}
The time evolution in the interaction picture is governed by the interaction Hamiltonian in Eq.(\ref{S10}), leading to the final state
\begin{equation}
|\Psi_f\rangle = \mathcal{T} \exp\left[ -i \int_{\mathbb{R}} dt \left( \frac{d\tau_A}{dt} H_A(\tau_A) + \frac{d\tau_B}{dt} H_B(\tau_B) \right) \right] |\Psi_i\rangle,
\end{equation}
where  $\mathcal{T}$ denotes time ordering with respect to the coordinate time $t$, with respect to which the field vacuum is defined.
Tracing over the field degrees of freedom, the reduced density matrix of the detectors, to leading order in perturbation theory \cite{SDF65,SDF68}, takes the form
\begin{equation}\label{S41}
\rho_{AB} := \mathrm{Tr}_\phi \left( \rho_{tot} \right) \\
= \begin{pmatrix}
1 - P_A - P_B & 0 & 0 & X \\
0 & P_B & C & 0 \\
0 & C^* & P_A & 0 \\
X^* & 0 & 0 & 0
\end{pmatrix} + \mathcal{O}(\lambda^4),
\end{equation}
written in the ordered basis $\{|0_A 0_B\rangle, |0_A 1_B\rangle\, |1_A 0_B\rangle, |1_A 1_B\rangle\}$.
The excitation probabilities and correlation terms are given by
\begin{align}
P_{D} &= \lambda^2 \int_{\mathbb{R}} d\tau \int_{\mathbb{R}} d\tau' \chi(\tau) \chi(\tau') e^{-i\Omega_D(\tau - \tau')} \times W_{\mathrm{BTZ}}^{\mathrm{Lv}}\left[ \mathbf{x}_D(\tau), \mathbf{x}_D(\tau') \right], \\
C &= \lambda^2 \int_{\mathbb{R}} d\tau \int_{\mathbb{R}} d\tau' \chi(\tau) \chi(\tau') e^{-i(\Omega_A\tau - \Omega_B\tau')} \times W_{\mathrm{BTZ}}^{\mathrm{Lv}}\left[ \mathbf{x}_A(\tau), \mathbf{x}_B(\tau') \right], \\
X &= -\lambda^2 \int_{\mathbb{R}} d\tau_A \int_{\mathbb{R}} d\tau_B \chi_A(\tau_A) \chi_B(\tau_B) e^{-i\Omega(\tau_A + \tau_B)} \notag \\
&\quad \times \left\{ \Theta\left[ t(\tau_A) - t(\tau_B) \right] W_{\mathrm{BTZ}}^{\mathrm{Lv}}\left[ \mathbf{x}_A(\tau_A), \mathbf{x}_B(\tau_B) \right] \right. \left. + \Theta\left[ t(\tau_B) - t(\tau_A) \right] W_{\mathrm{BTZ}}^{\mathrm{Lv}}\left[ \mathbf{x}_B(\tau_B), \mathbf{x}_A(\tau_A) \right] \right\}.
\end{align}
where $\Theta(t)$ is the Heaviside step function.  The off-diagonal terms $C$ and $X$ encode nonlocal correlations induced by the field.

We work within the Einstein-bumblebee gravity framework, which constitutes a Lorentz-violating extension of general relativity. We consider a $(2+1)$-dimensional static and circularly symmetric spacetime, whose line element is assumed to take the form  \cite{SY5}
\begin{eqnarray}\label{S1}
ds^{2}&=&-A(r)dt^{2}+F(r)dr^{2}+r^{2}d\phi^{2}.
\end{eqnarray}
where $A(r)$ and $F(r)$ are functions of the radial coordinate to be determined. The background bumblebee field $B_{\mu}$ is chosen as
\begin{eqnarray}\label{S2}
B_{\mu}=(0,b\sqrt{F(r)},0),
\end{eqnarray}
which satisfies the fixed-norm condition $B^{\mu}B_{\mu}=b^2$. With this ansatz, the nonvanishing components of the effective gravitational field equations and the equation of motion for the bumblebee field read
\begin{align}
\mathcal{G}_{tt} &= \frac{(1 + \alpha)A\partial_rF}{2rF^2} - A\Lambda, \label{S3} \\
\mathcal{G}_{rr} &= (\Lambda - b^2\kappa\lambda)F + \frac{(1 + \alpha)\partial_rA}{2rA} + \frac{\alpha\partial_rF}{2rF} + \frac{\alpha\Upsilon}{2A}, \label{S4} \\
\mathcal{G}_{\phi\phi} &= r^2\Lambda - \frac{(1 + \alpha)r^2\Upsilon}{2AF}, \label{S5} \\
\Pi_r &= \frac{1}{\kappa b\sqrt{F}} \left( \frac{\alpha\partial_rF}{2rF^2} + \frac{\alpha\Upsilon}{2AF} - \kappa b^2\lambda \right), \label{S6}
\end{align}
where $\alpha=\varrho b^{2}$ denotes the Lorentz-violating parameter, and
\begin{equation}
\Upsilon = \frac{(\partial_r A)^2}{2A} + \frac{\partial_r A\partial_r F}{2F} - \partial^2_r A.
\end{equation}
To solve the coupled differential Eqs.(\ref{S3})-(\ref{S6}), we consider the following linear combinations as
\begin{equation}
2rF^2\left(\mathcal{G}_{tt} + \frac{A}{F}\mathcal{G}_{rr} - \kappa b \sqrt{F} A \Pi_r\right) = 0, \quad \alpha \mathcal{G}_{\phi\phi} + \frac{(1+\alpha)r^2}{F} \mathcal{G}_{rr} = 0.
\end{equation}
The first relation immediately yields
\begin{equation}
\partial_r (AF) = 0 \implies F = C_1 / A,
\end{equation}
where $C_1$ is an integration constant. Substituting this result into the second relation, one obtains
\begin{equation}\label{S7}
\frac{(1 + \alpha)}{2C_1} r \partial_r A + r^2 \left[ \alpha \Lambda + (1 + \alpha)(\Lambda - \kappa b^2 \lambda) \right] = 0.
\end{equation}
Integrating Eq.(\ref{S7}), the metric function $A(r)$ is found to be
\begin{equation}
A(r) = r^2 C_1 \left( \kappa b^2 \lambda - \frac{1 + 2\alpha}{1 + \alpha} \Lambda \right) + C_2,
\end{equation}
with $C_2$ another integration constant. At this stage, Eqs.(\ref{S3})-(\ref{S6}) share a common factor
$(1+\alpha)\kappa b^2 \lambda-2\alpha \Lambda$.
Requiring all constraint equations to be satisfied leads to the relation $\lambda = \frac{2\alpha}{(1+\alpha)\kappa b^2}$. Consequently, the metric function reduces to
\begin{equation}
A(r) = -\frac{C_1 \Lambda}{1 + \alpha} r^2 + C_2.
\end{equation}
To recover the asymptotic behavior of a BTZ black hole, we set $C_1 = (1+\alpha)$ and assume a negative cosmological constant. Defining the AdS radius  $\ell = \sqrt{-1/\Lambda}$ and identifying $C_2=-M$, the resulting Lorentz-violating BTZ metric reads
\begin{equation}\label{S8}
{g_{\mu\nu} = \mathrm{diag}\left\{ M - \frac{r^2}{\ell^2}, \frac{1 + \alpha}{r^2/\ell^2-M}, r^2 \right\}.}
\end{equation}
The Kretschmann scalar is given by
$R^{\mu\nu\sigma\tau} R_{\mu\nu\sigma\tau} = \frac{12}{\ell^4 (1 + \alpha)^2}$,
indicating a curvature singularity at $\alpha=-1$ \cite{B7}.
The event horizon is located at $r_h = \ell\sqrt{M}$, which coincides with that of the standard BTZ black hole and is independent of the Lorentz-violating parameter.

Setting $C_2=0$, the spacetime reduces to a Lorentz-violating $AdS_3$ background with metric
\begin{equation}\label{S9}
\eta_{\mu\nu}^{\mathrm{Lv}} = \mathrm{diag}\left\{ -\frac{r^2}{\ell^2}, \frac{(1+\alpha)\ell^2}{r^2}, r^2 \right\}.
\end{equation}
For $\alpha=0$, the standard $AdS_3$ geometry is recovered. Equivalently, in Eqs.(\ref{S9}) can be obtained from the conventional $AdS_3$ spacetime through the rescalings
\begin{equation}
t \to \sqrt{1+\alpha}\, t, \quad \ell \to \sqrt{1+\alpha}\, \ell.
\end{equation}
We characterize detector correlations using the proper radial distance.  For $r_2 > r_1 > r_h$, the proper distance between $(t, r_1, \phi)$ and $(t, r_2, \phi)$ is
\begin{equation}
d(r_1, r_2) = \int_{r_1}^{r_2} \sqrt{g_{\mu\nu} d x^\mu d x^\nu} = \ell \sqrt{1+\alpha} \ln \left( \frac{r_2 + \sqrt{r_2^2 - r_h^2}}{r_1 + \sqrt{r_1^2 - r_h^2}} \right),
\end{equation}
We assume  $r_B > r_A > r_h$ and fix the detector separation
$d_{AB} := d(r_A, r_B)$.
The distances from the horizon are denoted by
$d_j := d(r_h, r_j)$, $j \in \{A, B\}$.

The Hawking temperature of the Lorentz-violating BTZ black hole is
\begin{equation}
T_H = \frac{r_h}{2\pi \ell^2 \sqrt{1+\alpha}},
\end{equation}
which decreases monotonically with increasing $\alpha$.
The local temperature measured at $r=r_j$ is given by
$T_j = \frac{T_H}{\gamma_j}$, where the redshift factor reads
\begin{equation}
\gamma_j = \sqrt{-g_{tt}} = \frac{1}{\ell} \sqrt{r_j^2 - r_h^2} \quad  (r_j \geq r_h).
\end{equation}
For $r_B > r_A > r_h$, the redshift factors for detectors Alice and Bob can be written in terms of the proper distances:
\begin{equation}
\gamma_{\mathrm{A}} = \frac{r_h}{\ell} \sinh \frac{d_{\mathrm{A}}}{\ell \sqrt{1+\alpha}}, \quad \gamma_{\mathrm{B}} = \frac{r_h}{\ell} \sinh \frac{d_{\mathrm{AB}} + d_{\mathrm{A}}}{\ell \sqrt{1+\alpha}}.
\end{equation}

We now consider a conformally coupled scalar field in the Lorentz-violating BTZ spacetime. As shown explicitly in Appendix~\ref{AppQuotient}, the Lorentz-violating BTZ metric can be written as a standard nonrotating BTZ metric with redefined parameters, while the angular identification $\phi\sim\phi+2\pi$ remains unchanged. It therefore retains the quotient structure of an open region of $AdS_3$ by a discrete isometry, which justifies the image construction below \cite{BanadosGeometry,LifschytzOrtiz}. Choosing the
Hartle-Hawking vacuum $|0\rangle$, the Wightman function can be constructed via the image method applied to the Lorentz-violating $AdS_3$ background \cite{SDF65}. Denoting the angular identification by
$\Gamma : (t,r,\phi) \rightarrow (t,r,\phi+2\pi)$ \cite{SDF53},
the Wightman function takes the form
\begin{equation}\label{WBTZimage}
W_{\mathrm{BTZ}}^{\mathrm{Lv}}(\mathbf{x},\mathbf{x}')
=
\sum_{n=-\infty}^{\infty}
W_{AdS_3}^{\mathrm{Lv}}(\mathbf{x},\Gamma^n\mathbf{x}')
=
\frac{1}{4\pi\ell\sqrt{2(1+\alpha)}}
\sum_{n=-\infty}^{\infty}
\Pi(\mathbf{x},\Gamma^n\mathbf{x}'),
\end{equation}
where
\begin{equation}
\Pi(\mathbf{x},\Gamma^n\mathbf{x}')
=
\frac{1}{\sqrt{\sigma_n}}
-
\frac{\zeta}{\sqrt{\sigma_n+2}},
\end{equation}
\begin{equation}
\sigma_n
=
\frac{rr'}{r_h^2}
\cosh\!\left[
\frac{r_h(\Delta\phi-2\pi n)}{\ell\sqrt{1+\alpha}}
\right]
-
\frac{\sqrt{(r^2-r_h^2)(r'^2-r_h^2)}}{r_h^2}
\cosh\!\left(
\frac{r_h\Delta t}{\ell^2\sqrt{1+\alpha}}
\right)
-1 .
\end{equation}
Here, $\Delta\phi=\phi-\phi'$ and $\Delta t=t-t'$.
The parameter $\zeta\in\{-1,0,1\}$  specifies Neumann, transparent, and Dirichlet boundary conditions, respectively. Different choices of $\zeta$ correspond to different boundary configurations and modify the image contributions in the Wightman function, thereby affecting the quantitative behavior of the harvested quantum steering. In particular, Neumann and transparent boundary conditions can lead to different interference patterns between the direct and image terms of the field correlations. In this work, we focus on the Dirichlet boundary condition ($\zeta=1$), which represents a perfectly reflecting boundary configuration. This choice allows us to investigate more clearly the influence of Lorentz violation and gravitational redshift on the directional asymmetry of quantum steering without introducing additional boundary-related parameters. Although different boundary conditions may quantitatively modify the magnitude and operational range of the harvested steering, we expect the underlying mechanism responsible for the steering asymmetry, namely the inequivalence of local environments induced by gravitational redshift, to remain qualitatively valid. A systematic comparison among different boundary conditions is an interesting extension of the present study.

In the Lorentz-violating BTZ background, all quantities $P_A$, $P_B$, $C$, and $X$ depend explicitly on the Lorentz-violating parameter $\alpha$ through both the modified Wightman function and the detector trajectories. In what follows, we restrict to two identical static detectors located along the same radial direction ($\Delta\phi = 0$), with identical Gaussian switching functions activated synchronously. Under these assumptions, performing an appropriate change of variables yields a form of $C$ suitable for numerical evaluation \cite{SDF53},
\begin{align}
\!\!\!C &\!=\! \lambda^2 \int_{\mathbb{R}} d\tau_{\mathrm{A}} \int_{\mathbb{R}} d\tau_{\mathrm{B}} e^{-\tau_{\mathrm{A}}^2/2\sigma^2} e^{-\tau_{\mathrm{B}}^2/2\sigma^2} e^{i\Omega(\tau_{\mathrm{A}} - \tau_{\mathrm{B}})} W_{\mathrm{BTZ}}^{\mathrm{Lv}}\left[ \mathbf{x}_{\mathrm{A}}(\tau_{\mathrm{A}}), \mathbf{x}_{\mathrm{B}}(\tau_{\mathrm{B}}) \right] \nonumber \\
&\!=\! 2K\!\! \sum_{n=-\infty}^{\infty} \!\mathrm{Re}\!\! \int_{0}^{\infty} dx e^{-ax^2}\! e^{-i\beta x} \left[ (\cosh \chi_{\mathrm{AB},n}^- \!-\! \cosh x)^{-1/2} \!-\! \zeta (\cosh \chi_{\mathrm{AB},n}^+ \!-\! \cosh x)^{-1/2} \right],
\end{align}
where $\alpha, \beta, K$ and $\chi^{\pm}_{AB,n}$ are defined by
\begin{equation}
a := \frac{\gamma_{\mathrm{A}}^2 \gamma_{\mathrm{B}}^2}{2\sigma^2(\gamma_{\mathrm{A}}^2 + \gamma_{\mathrm{B}}^2)} \frac{\ell^4(1 + \alpha)}{r_h^2}, \notag \\
\beta := \frac{\gamma_{\mathrm{A}} \gamma_{\mathrm{B}} (\gamma_{\mathrm{A}} + \gamma_{\mathrm{B}})}{\gamma_{\mathrm{A}}^2 + \gamma_{\mathrm{B}}^2} \frac{\ell^2 \sqrt{1+\alpha}}{r_h} \Omega, \notag \\
\end{equation}
\begin{equation}
K := \frac{\lambda^2 \sigma}{4} \sqrt{\frac{\gamma_{\mathrm{A}} \gamma_{\mathrm{B}}}{\pi(\gamma_{\mathrm{A}}^2 + \gamma_{\mathrm{B}}^2)}} \exp\left[ -\frac{\Omega^2 \sigma^2 (\gamma_{\mathrm{A}} - \gamma_{\mathrm{B}})^2}{2(\gamma_{\mathrm{A}}^2 + \gamma_{\mathrm{B}}^2)} \right], \notag \\
\end{equation}
\begin{equation}
\chi_{\mathrm{AB},n}^\pm := \mathrm{arccosh}\left[ \frac{r_h^2}{\ell^2 \gamma_{\mathrm{A}} \gamma_{\mathrm{B}}} \left( \frac{r_{\mathrm{A}} r_{\mathrm{B}}}{r_h^2} \cosh\left[ \frac{2\pi n r_h}{\ell \sqrt{1+\alpha}} \right] \pm 1 \right) \right]. \notag \\
\end{equation}

Similarly, one obtains numerical expressions for $P_A$, $P_B$, and $X$ in the Lorentz-violating background. Since $P_A$ and $P_B$ take the same form, we denote them collectively by $P_D$, given by
\begin{align}
 P_D&= -\frac{\zeta \lambda^2 \sigma}{2\sqrt{2\pi}} \mathrm{Re} \int_{0}^{\infty} dx \frac{e^{-a_{\mathrm{D}} x^2} e^{-i\beta_{\mathrm{D}} x}}{\sqrt{\cosh \chi_{\mathrm{D},0}^+ - \cosh x}}  \notag\\
&\quad + \frac{\lambda^2 \sigma^2}{2} \int_{\mathbb{R}} dx \frac{e^{-\sigma^2 (x - \Omega)^2}}{e^{x/T_{\mathrm{D}}} + 1} + \frac{\lambda^2 \sigma}{\sqrt{2\pi}} \sum_{n=1}^{\infty} \mathrm{Re} \int_{0}^{\infty} dx  \notag\\
&\quad \times e^{-a_{\mathrm{D}} x^2} e^{-i\beta_{\mathrm{D}} x} \left( \frac{1}{\sqrt{\cosh \chi_{\mathrm{D},n}^- - \cosh x}} - \frac{\zeta}{\sqrt{\cosh \chi_{\mathrm{D},n}^+ - \cosh x}} \right),
\end{align}
where
\begin{align}
a_{\mathrm{D}} &:= \frac{\ell^4 \gamma_{\mathrm{D}}^2 (1 + \alpha)}{4\sigma^2 r_h^2}, \quad \beta_{\mathrm{D}} := \frac{\sqrt{1+\alpha} \ell^2 \gamma_{\mathrm{D}} \Omega}{r_h},  \notag \\
\chi_{\mathrm{D},n}^\pm &:= \mathrm{arccosh}\left[ \frac{r_h^2}{\ell^2 \gamma_{\mathrm{D}}^2} \left( \frac{r_{\mathrm{D}}^2}{r_h^2} \cosh\left[ \frac{2\pi n r_h}{\ell \sqrt{1+\alpha}} \right] \pm 1 \right) \right]. \notag
\end{align}
The first two terms, corresponding to $(n = 0)$, resemble AdS-Rindler contributions in a Lorentz-violating spacetime, whereas the last term $(n \neq 0)$ is known as the BTZ term. We also obtain the matrix element $X$
\begin{align}
 X &= -\sum_{n=-\infty}^{\infty} \left[ K_{\mathrm{M}} \int_{0}^{\infty} dx \frac{\exp(-a_{\mathrm{M}} x^2) \cos(\beta_{\mathrm{M}} x)}{\sqrt{\cosh \chi_{\mathrm{M},n}^- - \cosh x}} \right.  \notag\\
&\quad \left. -\zeta K_{\mathrm{M}} \int_{0}^{\infty} dx \frac{\exp(-a_{\mathrm{M}} x^2) \cos(\beta_{\mathrm{M}} x)}{\sqrt{\cosh \chi_{\mathrm{M},n}^+ - \cosh x}} \right],
\end{align}
with corresponding definitions for
\begin{align}
K_{\mathrm{M}} &:= \frac{2\sqrt{\pi}}{\pi} K, \quad \chi_{\mathrm{M},n}^\pm := \chi_{\mathrm{AB},n}^\pm, \quad a_{\mathrm{M}} := a,  \notag \\
\beta_{\mathrm{M}} &:= \frac{\gamma_{\mathrm{A}} \gamma_{\mathrm{B}} (\gamma_{\mathrm{A}} - \gamma_{\mathrm{B}})}{\gamma_{\mathrm{A}}^2 + \gamma_{\mathrm{B}}^2} \frac{\ell^2 \sqrt{1+\alpha}}{r_h} \Omega. \notag
\end{align}

In the numerical evaluation of the above integrals, we take the complex square root on its principal branch, $-\pi<\arg z\leq\pi$. Consistently with the Wightman $i\epsilon$ prescription, the integration contour approaches the positive real axis from the lower half-plane, $x\to x-i0$ \cite{SDF53}. Thus,
\begin{equation*}
\cosh\chi-\cosh(x-i0)
=\cosh\chi-\cosh x+i0\,\sinh x,
\end{equation*}
and, for $q>0$, $\sqrt{-q+i0}=i\sqrt q$. Equivalently,
\begin{equation*}
\frac{1}{\sqrt{\cosh\chi-\cosh(x-i0)}}=
\begin{cases}
\displaystyle \frac{1}{\sqrt{\cosh\chi-\cosh x}}, & 0\leq x<\chi,\\[7pt]
\displaystyle -\frac{i}{\sqrt{\cosh x-\cosh\chi}}, & x>\chi.
\end{cases}
\end{equation*}
In practice, each integral is split at the corresponding branch point $x=\chi$, where the square-root singularity is integrable. The same prescription is used for every square-root factor in the expressions for $C$, $P_D$, and $X$.

In the Lorentz-violating BTZ background, the parameter $\alpha$ deforms both the geometry and the field correlations, entering through the modified Wightman function and the redshift factors along the detector trajectories. The evaluation of $C$, $P_D$, and $X$ follows a common strategy: the double integrals are reduced via suitable coordinate transformations, together with the corresponding Jacobians, leading to the above single-integral representations. As a result, $\alpha$ modifies the effective temperature and correlation structure, yielding deviations from the standard BTZ case.

\section{Quantum steering harvested and its asymmetry near Lorentz-violating BTZ black
hole}
In this section, we analyze the impact of Lorentz-violating on the harvested steering, including its asymmetry, magnitude, and range. Using Eqs.(\ref{S25}), (\ref{S26}) and (\ref{S41}), the steering measures  $S^{A \rightarrow B}$ and $S^{B \rightarrow A}$ can be written as
\begin{equation}\label{QB1}
S^{B \to A} \!=\! \max\left\{ 0, |X| -\! \sqrt{\tfrac{1+\sqrt{3}}{2} P_A P_B \!+\! \tfrac{1}{2} P_A \!-\! \tfrac{1}{2} P_A^2}, \right. \\
\left. |C| \!-\! \sqrt{\tfrac{1-\sqrt{3}}{2} P_A P_B \!+\! \tfrac{1}{2} P_A \!-\! \tfrac{1}{2} P_A^2} \right\},
\end{equation}
\begin{equation}\label{QB2}
S^{A \to B} \!=\! \max\left\{ 0, |X| \!-\! \sqrt{\tfrac{1+\sqrt{3}}{2} P_A P_B + \tfrac{1}{2} P_B \!-\! \tfrac{1}{2} P_B^2}, \right. \\
\left. |C| \!-\! \sqrt{\tfrac{1-\sqrt{3}}{2} P_A P_B + \tfrac{1}{2} P_B \!-\! \tfrac{1}{2} P_B^2} \right\}.
\end{equation}

\begin{figure}
\centering
\includegraphics[height=2.6in,width=5.5in]{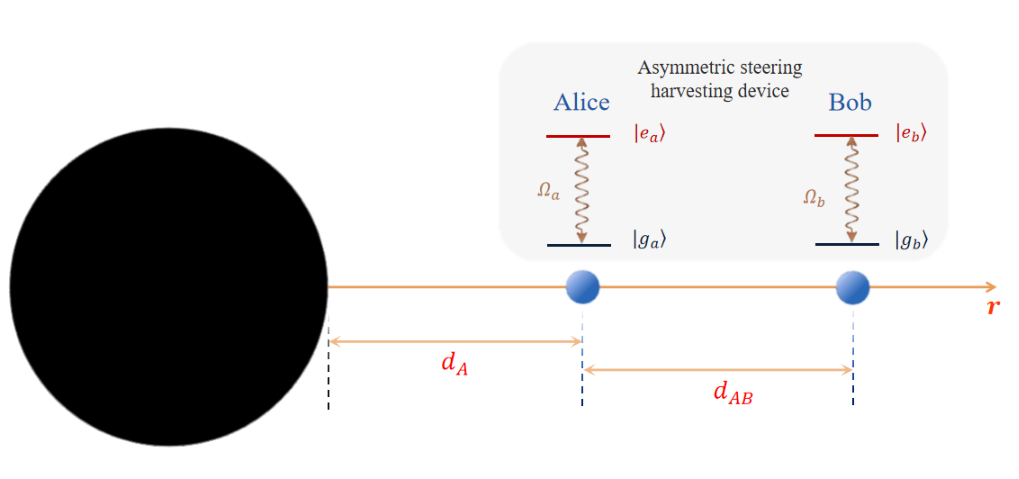}
\caption{Schematic diagram of the Unruh-DeWitt detector in an asymmetric quantum steering harvesting device. Static detectors Alice and Bob are placed on the same side of the Lorentz-violating black hole.}
\label{Fig.1}
\end{figure}

In Fig.\ref{Fig.1}, we illustrate the spatial configuration of the detectors relative to the black hole. It is manifest from the setup that the quantum steering exhibits an inherent asymmetry. To quantify this directional asymmetry, we define the steering asymmetry as
\begin{equation}
S_{AB}^\Delta = | S^{B \to A} - S^{A \to B} |.
\end{equation}
A nonvanishing value of $S_{AB}^{\Delta}$ signals inequivalent steerabilities between the two directions, i.e., the ability of $A$ to steer $B$ differs from that of $B$ to steer $A$. This asymmetry originates from the imbalance in the local noise and correlations experienced by the detectors, encoded in the unequal excitation probabilities ($P_A \neq P_B$) and the trajectory-dependent field correlations. In particular, $S_{AB}^{\Delta}=0$ corresponds to symmetric steering, while $S_{AB}^{\Delta}>0$ indicates directional bias. In the present setup, such asymmetry is further influenced by the Lorentz-violating BTZ black
hole. It is instructive to consider the limit of identical detector environments. When $d_{AB}\rightarrow0$, the two detectors experience the same gravitational redshift, yielding $P_A=P_B$. Therefore, the two directional steering quantities become identical and $S_{AB}^\Delta=0$. The finite asymmetry observed for separated detectors is thus a direct consequence of the gravitationally induced difference between their local environments. For completeness, the analytical behaviors of the directional quantum steering measures in several physically relevant limiting regimes, including $d_{AB}\rightarrow0$, $d_{AB}\rightarrow\infty$, and $d_A\rightarrow\infty$, are discussed in Appendix~\ref{AppQuotientB}.

\begin{figure}
\begin{minipage}[t]{0.35\linewidth}
\centering
\includegraphics[width=2.0in,height=4cm]{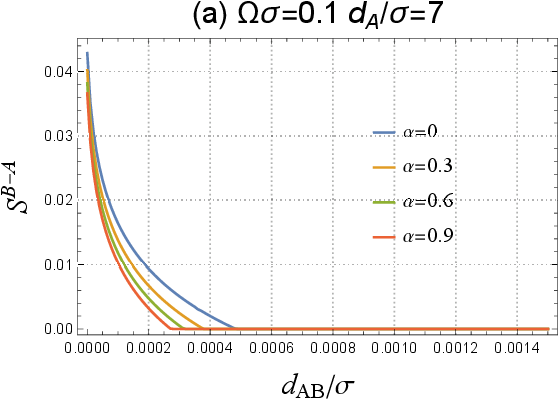}
\label{fig1a}
\end{minipage}%
\begin{minipage}[t]{0.35\linewidth}
\centering
\includegraphics[width=2.0in,height=4cm]{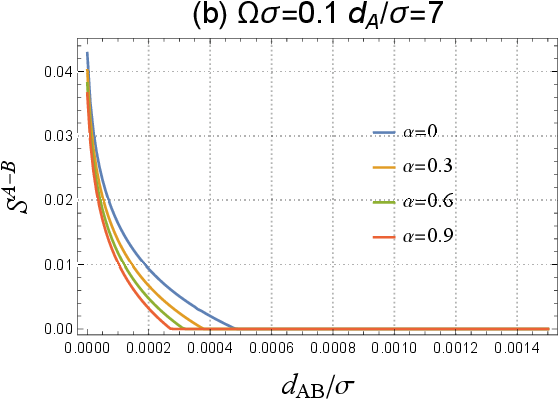}
\label{fig1b}
\end{minipage}%
\begin{minipage}[t]{0.35\linewidth}
\centering
\includegraphics[width=2.0in,height=4cm]{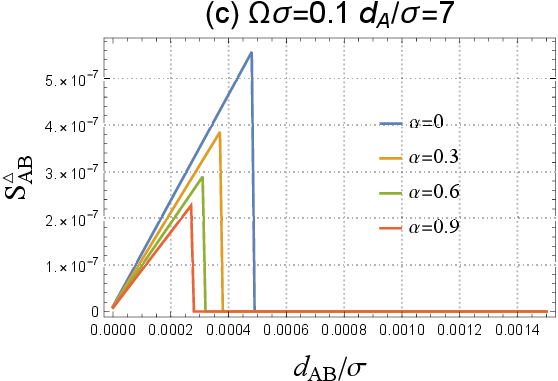}
\label{fig1c}
\end{minipage}%

\caption{Quantum steering $S^{A\rightarrow B}$, $S^{B\rightarrow A}$, and the steering asymmetry $S^{\Delta}_{AB}$ between two detectors as a function of the detector separation $d_{AB}/\sigma$ with $\Omega\sigma = 0.1$ and $d_{A}/\sigma = 7$. For all numerical calculations, the infinite summation over is set to $n_{\max}=5$.}
\label{Fig.2}
\end{figure}

In Fig.\ref{Fig.2}, we display the harvested quantum steering $S^{A\to B}$ and $S^{B\to A}$, together with the steering asymmetry $S^{\Delta}_{AB}$, as a function of the detector separation $d_{AB}/\sigma$. The results are shown for a fixed energy gap $\Omega\sigma = 0.1$. The numerical calculations are performed with the image-sum truncation $(n_{\max}=5)$ and the numerical integration parameters are specified in Appendix~\ref{AppQuotientC}. For fixed detector parameters, the magnitude of steering decreases monotonically with increasing separation and vanishes beyond a critical distance, signaling the sudden death of quantum steering. A clear directional asymmetry,  $S^{A\to B} > S^{B\to A}$, is observed. This behavior originates from the inequivalent local environments of the detectors: due to gravitational redshift, Alice experiences stronger effective thermal noise than Bob. Since steering is constrained by the noise affecting the steered party, Bob's lower-noise environment renders him more susceptible to being steered, thereby enhancing $S^{A\to B}$ relative to $S^{B\to A}$. In the limit of vanishing separation, the detectors experience identical gravitational conditions and the steering becomes symmetric. As the separation increases, the mismatch in local redshift factors grows, leading to an enhancement of the asymmetry. However, beyond a certain separation scale, the asymmetry rapidly disappears, indicating a non-monotonic dependence on the detector separation. Finally, by comparison with the standard BTZ case $(\alpha=0)$, we find that Lorentz-violating generically suppresses the harvesting of quantum steering. This suppression manifests in two aspects: the overall magnitude of the harvested steering is reduced, and the operational regime is narrowed, as reflected by a shorter harvesting range and a diminished steering asymmetry.

\begin{figure}
\begin{minipage}[t]{0.35\linewidth}
\centering
\includegraphics[width=2.0in,height=4cm]{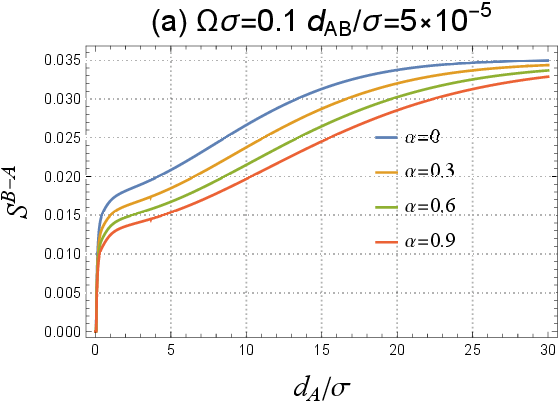}
\label{fig2a}
\end{minipage}%
\begin{minipage}[t]{0.35\linewidth}
\centering
\includegraphics[width=2.0in,height=4cm]{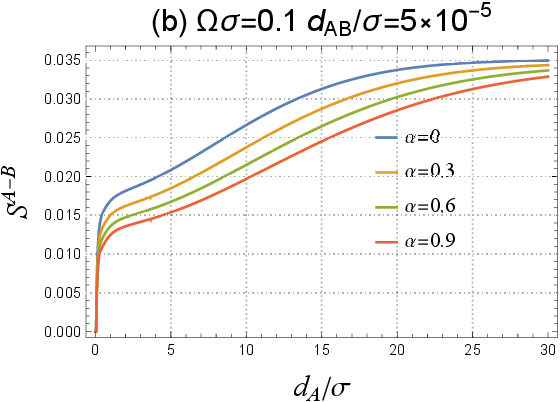}
\label{fig2b}
\end{minipage}%
\begin{minipage}[t]{0.35\linewidth}
\centering
\includegraphics[width=2.0in,height=4cm]{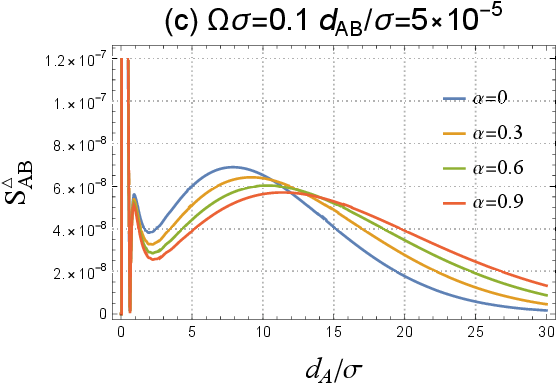}
\label{fig2c}
\end{minipage}%

\caption{Quantum steering $S^{A\rightarrow B}$, $S^{B\rightarrow A}$, and the steering asymmetry $S^{\Delta}_{AB}$ between two detectors as a function of the distance detector Alice from the horizon $d_{A}/\sigma$ with $\Omega\sigma = 0.1$, $d_{AB}/\sigma = 5\times10^{-5}$ and \textcolor{red}{$n_{\max}=5$}.}
\label{Fig.3}
\end{figure}

In Fig.\ref{Fig.3}, we present the harvested quantum steering as a function of the distance detector Alice from the horizon $d_{A}/\sigma$  for several values of the Lorentz-violating parameter $\alpha$, while keeping the energy gap fixed. As detector $A$ approaches the horizon, the harvested steering is strongly suppressed. This behavior can be attributed to the rapid increase of the local Hawking temperature, which enhances thermal noise and degrades quantum correlations. Consequently, the ability to extract steering from the field becomes progressively weaker in the near-horizon region. A more intricate behavior is observed in the steering asymmetry. As shown in Fig.\ref{Fig.2}(c), the difference between the two directional steerabilities exhibits pronounced fluctuations close to the horizon, reflecting the strong sensitivity of steering to local redshift effects. In contrast, as the detectors move away from the horizon, the local environments become increasingly similar, and the two directional steerabilities gradually converge, leading to a suppression of asymmetry.  Finally, for a fixed energy gap, increasing the radial distance of detector $A$ enlarges the parameter region in which nonvanishing steering can be observed. This indicates that moving the detector away from the high-temperature near-horizon region is beneficial for the harvesting of quantum steering.

\begin{figure}
\begin{minipage}[t]{0.35\linewidth}
\centering
\includegraphics[width=2.0in,height=4cm]{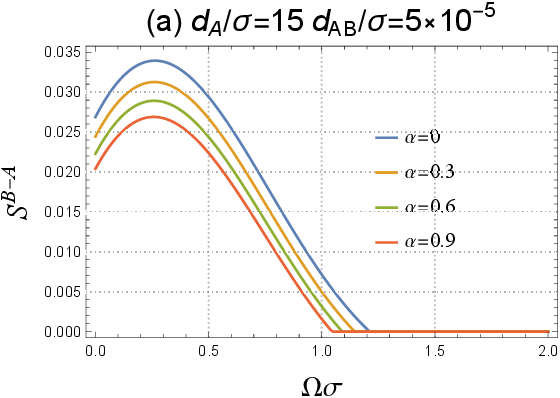}
\label{fig3a}
\end{minipage}%
\begin{minipage}[t]{0.35\linewidth}
\centering
\includegraphics[width=2.0in,height=4cm]{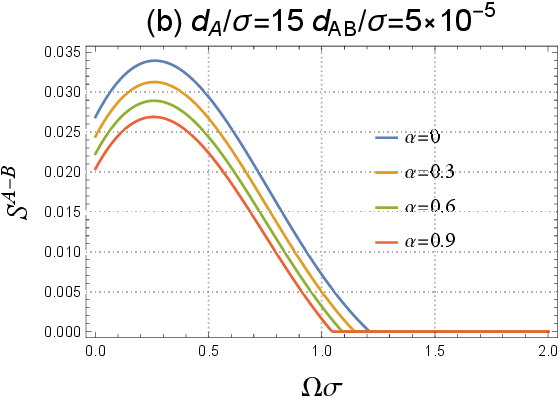}
\label{fig3b}
\end{minipage}%
\begin{minipage}[t]{0.35\linewidth}
\centering
\includegraphics[width=2.0in,height=4cm]{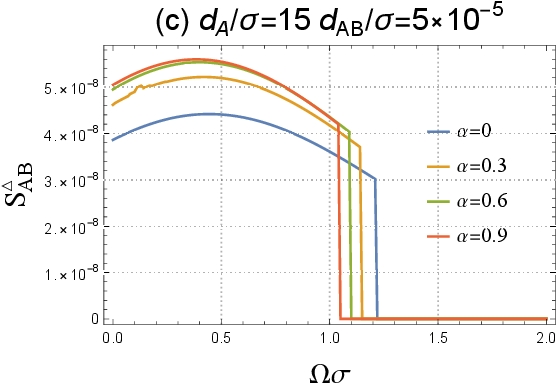}
\label{fig3c}
\end{minipage}%

\caption{Quantum steering $S^{A\rightarrow B}$, $S^{B\rightarrow A}$, and the steering asymmetry $S^{\Delta}_{AB}$ between two detectors as a function of the energy gap $\Omega\sigma$ with $d_{A}/\sigma = 15$, $d_{AB}/\sigma = 5\times10^{-5}$ and \textcolor{red}{$n_{\max}=5$.}}
\label{Fig.4}
\end{figure}

In Fig.\ref{Fig.4}, we show the harvested quantum steering as a function of the detector energy gap $\Omega\sigma$ for several values of the Lorentz-violating parameter $\alpha$, with detector $A$ placed at a fixed radial position. For a given detector configuration, steering is nonvanishing only within a finite interval of $\Omega\sigma$. Outside this interval, the harvested steering undergoes sudden death, indicating that the extraction of nonclassical correlations is possible only in a restricted energy window. Within this window, the steering exhibits a clear maximum at an optimal energy gap $\Omega_{\mathrm{opt}}$, which sets the most efficient energy scale for harvesting. This behavior reflects the interplay between the detector energy scale and the characteristic frequencies of the field correlations. The effect of Lorentz violation is twofold. First, increasing $\alpha$ suppresses the overall magnitude of the harvested steering, with the reduction being most pronounced near $\Omega_{\mathrm{opt}}$. Second, it narrows the admissible energy window, thereby reducing the parameter region where steering can be sustained. Overall, these results indicate that both the efficiency and robustness of quantum steering harvesting are highly sensitive to the detector energy scale, and that Lorentz violation further limits the accessible regime for extracting directional quantum correlations in the BTZ background.

\section{Conclusions}
In this work, we have investigated the harvesting of quantum steering and its directional asymmetry using two static Unruh-DeWitt detectors coupled to a scalar field in a Lorentz-violating BTZ black hole spacetime. By deriving analytical expressions for quantum steering in both directions within this framework, we have systematically analyzed how Lorentz violation reshapes field correlations and, consequently, the extraction of nonclassical resources. Our results reveal several key physical features: \textbf{(i) Universal suppression and contraction of the operational regime:}
Lorentz violation generically suppresses the harvesting of quantum steering. This effect manifests not only as a reduction in the overall steering strength, but also as a contraction of the operational parameter space, including both the maximal separation over which steering can be sustained and the admissible range of detector energy gaps; \textbf{(ii) Redshift-induced asymmetry and noise-governed steerability:}
The interplay between gravitational redshift and local thermal effects leads to a pronounced directional asymmetry, characterized by $S^{A\to B} > S^{B\to A}$. This asymmetry originates from the inequivalence of local environments: the detector subject to lower effective thermal noise is more susceptible to being steered. Remarkably, we uncover a counterintuitive feature, namely that the observer immersed in a higher effective thermal noise can exhibit stronger steerability, highlighting the nontrivial and intrinsically directional nature of quantum steering in curved spacetime;
\textbf{(iii) Finite energy window, sudden death, and optimal resonance:}
We show that quantum steering harvesting is only possible within a finite interval of the detector energy gap. Outside this interval, steering undergoes sudden death, indicating that the extraction of nonclassical correlations is restricted to a limited energy window. Within this viable region, an optimal energy gap $\Omega_{\text{opt}}$ maximizes the harvested steering, reflecting a resonance-like matching between the detector energy scale and the dominant frequencies of the field correlations. Notably, the suppressive effect of Lorentz violation is most pronounced near this optimal point, indicating enhanced sensitivity to symmetry breaking at maximal correlation extraction. Taken together, these results indicate that Lorentz violation acts as a fundamental geometric constraint on the quantum information capacity of spacetime. In the context of quantum steering, this constraint manifests not only as a suppression of correlation strength, but also as a reduction of its intrinsic directionality, thereby limiting the extent to which  quantum correlations can be asymmetrically distributed. From a broader perspective, this behavior can be understood as a redistribution of directional  correlation resources among competing quantum degrees of freedom in a Lorentz-violating geometry. Furthermore, the influence of alternative boundary configurations, such as Neumann and transparent boundary conditions, on the harvesting of directional quantum steering remains an interesting problem. A systematic investigation of how boundary-induced modifications of field correlations compete with Lorentz-violating effects will be left for future work.

\begin{acknowledgments}
This work is supported by the National Natural
Science Foundation of China (Grant Nos.12575056 and 12547147), the China Postdoctoral Science Foundation (Grant No. 2025M783393), the Natural
Science Foundation of the Liaoning Scientific Committee No. 2026-MS-281, and LiaoNing Revitalization Talents Program (XLYC2503099).	
\end{acknowledgments}

\appendix
\section{Quotient construction of the Lorentz-violating BTZ spacetime}\label{AppQuotient}

For $\alpha>-1$, define the effective AdS radius, mass parameter, and time coordinate by
\begin{equation}\label{A1}
L=\ell\sqrt{1+\alpha},\qquad
\widetilde M=\frac{M}{1+\alpha},\qquad
T=\sqrt{1+\alpha}\,t.
\end{equation}
The metric in Eq.~(\ref{S8}) then becomes
\begin{equation}\label{A2}
ds^2=-\left(\frac{r^2}{L^2}-\widetilde M\right)dT^2
+\frac{dr^2}{r^2/L^2-\widetilde M}+r^2d\phi^2,
\end{equation}
which is the standard nonrotating BTZ metric with AdS radius $L$ and mass parameter $\widetilde M$. Its horizon radius is unchanged,
$r_h=L\sqrt{\widetilde M}=\ell\sqrt{M}$. Since the transformation in Eq.~(\ref{A1}) leaves the angular coordinate unchanged, the identification $\phi\sim\phi+2\pi$ is preserved. Hence the Lorentz-violating BTZ geometry has the same quotient construction as the standard BTZ spacetime \cite{BanadosGeometry}:
\begin{equation}\label{A3}
\mathcal{M}_{\mathrm{BTZ}}^{\mathrm{Lv}}
=\mathcal{U}_{AdS_3(L)}/\langle\Gamma\rangle,
\qquad
\Gamma:(T,r,\phi)\mapsto(T,r,\phi+2\pi),
\end{equation}
where $\langle\Gamma\rangle\simeq\mathbb{Z}$ and $\mathcal{U}_{AdS_3(L)}$ is the corresponding open region of $AdS_3$ with radius $L$. In the $\mathbb{R}^{2,2}$ embedding, $\Gamma$ is the usual discrete boost with rapidity $2\pi r_h/L$. The bumblebee background is independent of $\phi$ and is therefore compatible with this identification.

The Wightman function can consequently be obtained from the standard BTZ image construction \cite{LifschytzOrtiz}:
\begin{equation}\label{A4}
W_{\mathrm{BTZ}}^{\mathrm{Lv}}(x,x')
=\sum_{n\in\mathbb{Z}}W_{AdS_3(L)}(x,\Gamma^n x').
\end{equation}
In the standard BTZ expression, the prefactor and the two dimensionless arguments are
\begin{equation}\label{A5}
\frac{1}{4\pi L\sqrt{2}},\qquad
\frac{r_h(\Delta\phi-2\pi n)}{L},\qquad
\frac{r_h\Delta T}{L^2}.
\end{equation}
Using Eq.~(\ref{A1}), these become
\begin{equation}\label{A6}
\frac{1}{4\pi\ell\sqrt{2(1+\alpha)}},\qquad
\frac{r_h(\Delta\phi-2\pi n)}{\ell\sqrt{1+\alpha}},\qquad
\frac{r_h\Delta t}{\ell^2\sqrt{1+\alpha}},
\end{equation}
respectively. They reproduce exactly the image sum and $\sigma_n$ in Eq.~(\ref{WBTZimage}), thereby justifying the use of the method of images. This equivalence concerns the background metric seen by the scalar field; the underlying bumblebee configuration still distinguishes the Lorentz-violating theory from ordinary Einstein gravity.

\section{Analytical limiting behaviors of quantum steering}\label{AppQuotientB}
It is instructive to examine the quantum steering measures in several physically relevant limiting regimes. First, when the separation between the two detectors approaches zero, \(d_{AB}\rightarrow0\), the two detectors experience the same local gravitational environment. In this limit, one has \(P_A=P_B\equiv P\), and Eqs.(\ref{QB1}) and (\ref{QB2})  reduce to
$$
S^{A\rightarrow B}=S^{B\rightarrow A}
=\max\left\{
0,\,
|X|-\sqrt{\frac{P}{2}+\frac{\sqrt{3}}{2}P^2},\,
|C|-\sqrt{\frac{P}{2}-\frac{\sqrt{3}}{2}P^2}
\right\}.
$$
Consequently, the directional steering becomes symmetric, and the steering asymmetry vanishes,
$$
S^\Delta_{AB}=|S^{A\rightarrow B}-S^{B\rightarrow A}|=0.
$$
This result is physically expected, since the two detectors are subject to identical local conditions in the coincident-detector limit.
Second, in the large-separation limit, \(d_{AB}\rightarrow\infty\), the field-mediated correlations between the two detectors vanish. In particular, the nonlocal correlation functions \(C\) and \(X\) approach zero,
$$
C,X\rightarrow0.
$$
Since the remaining terms in Eqs.(\ref{QB1}) and (\ref{QB2})  are non-negative, the two directional steering measures consequently satisfy
$$
S^{A\rightarrow B}\rightarrow0,
\qquad
S^{B\rightarrow A}\rightarrow0,
\qquad
d_{AB}\rightarrow\infty.
$$
Thus, quantum steering cannot be sustained when the two detectors are sufficiently far apart. This limiting behavior is consistent with the finite critical separation observed in Fig.\ref{Fig.2}. Finally, consider the limit in which the detectors are located far away from the black-hole horizon, \(d_A\rightarrow\infty\). The local redshift factors increase with the proper distance from the horizon, leading to
$$
T_A=\frac{T_H}{\gamma_A}\rightarrow0,
\qquad
T_B=\frac{T_H}{\gamma_B}\rightarrow0,
$$
where \(T_A\) and \(T_B\) are the local temperatures experienced by Alice and Bob, respectively. Therefore, the Hawking-induced thermal effects become negligible in this limit, and the detector dynamics approaches that of the zero-temperature regime. Accordingly, the steering measures approach their corresponding asymptotic values,
$$
S^{A\rightarrow B}\rightarrow S^{A\rightarrow B}_{T=0},
\qquad
S^{B\rightarrow A}\rightarrow S^{B\rightarrow A}_{T=0},
\qquad
d_A\rightarrow\infty.
$$
Hence, sufficiently far from the black-hole horizon, the suppression of quantum steering caused by the Hawking thermal environment disappears, and the residual steering is determined by the detector-field correlations in the zero-temperature limit.

\section{Numerical convergence and integration precision}\label{AppQuotientC}

\begin{figure}
\begin{minipage}[t]{0.5\linewidth}
\centering
\includegraphics[width=3in,height=5cm]{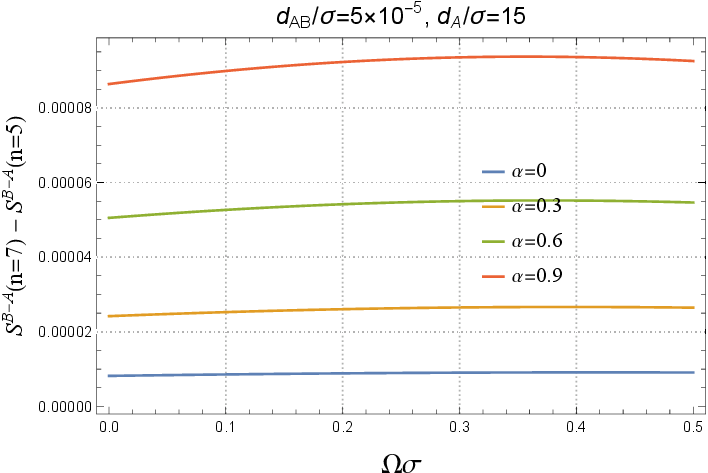}
\label{figc1}
\end{minipage}%
\begin{minipage}[t]{0.5\linewidth}
\centering
\includegraphics[width=3in,height=5
cm]{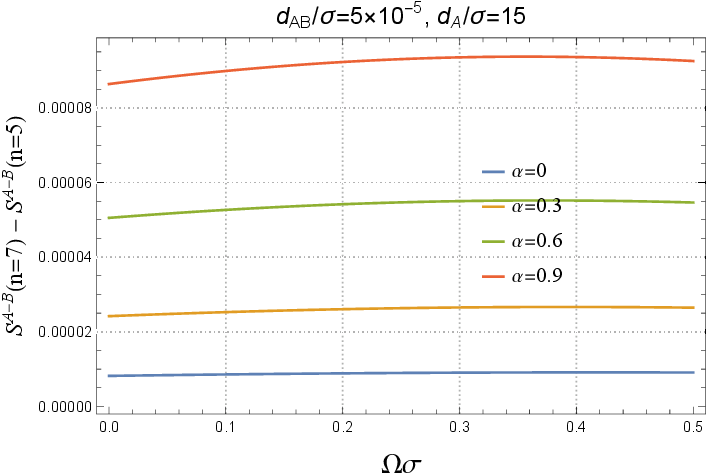}
\label{figc2}
\end{minipage}%

\caption{The absolute difference between the quantum steering measures calculated by truncating the infinite sum at $n_{\max}=7$ and $n_{\max}=5$, plotted as a function of the detector energy gap $\Omega\sigma$.}
\label{Fig.C1}
\end{figure}

To establish the numerical reliability of our results, we examine the convergence associated with the truncation of the infinite image sum and specify the numerical settings adopted for the integrations. In the calculations presented in this work, the infinite sum over $n$ is truncated at $n_{\max}=5$. To quantify the influence of this cutoff, we repeat the calculations with $n_{\max}=7$ and compare the resulting steering measures $S^{B\to A}$ and $S^{A\to B}$ with those obtained for $n_{\max}=5$. As shown in Fig.~\ref{Fig.C1}, the absolute difference between the results obtained with the two truncation cutoffs remains on the order of $10^{-5}$ across the parameter range considered, corresponding to a relative difference of approximately $0.3\%$. 
The smallness of these differences demonstrates that the contribution from higher-order image terms is already sufficiently suppressed at $n_{\max}=5$, so that the residual truncation error has no appreciable influence on either the numerical results or the physical conclusions of this work. We therefore adopt $n_{\max}=5$ in the numerical calculations, which provides a satisfactory balance between computational efficiency and numerical accuracy.

For the numerical integrations, we employ the double-exponential integration method with $\texttt{MaxRecursion}=20$, $\texttt{WorkingPrecision}=20$, $\texttt{PrecisionGoal}=4$, and $\texttt{AccuracyGoal}=4$. These numerical settings are applied uniformly in the evaluation of the transition probabilities and the field-correlation terms entering the steering measures. Together with the convergence test for the image-sum truncation, these choices provide a consistent control of the numerical accuracy and ensure that the results reported in Figs.~\ref{Fig.2}--\ref{Fig.4} are numerically stable within the precision relevant to the present analysis.


\end{document}